\documentclass[twocolumn,prl]{revtex4}
\usepackage{graphicx}
\usepackage{dcolumn}
\usepackage{bm}

\begin{document}
%


\title{Critical behavior in vacuum gravitational collapse in $4+1$
dimensions}

\author{Piotr Bizo\'n}
\affiliation{M. Smoluchowski Institute of Physics, Jagellonian
University, Reymonta 4, 30-059 Krak\'ow, Poland}
\author{Tadeusz Chmaj}
\affiliation{H. Niewodniczanski Institute of Nuclear
   Physics, Polish Academy of Sciences,  Krak\'ow, Poland}
\author{Bernd G. Schmidt}
    \affiliation{Max-Planck-Institut f\"ur Gravitationsphysik,
Albert-Einstein-Institut, Am M\"uhlenberg 1, D-14476 Golm,
Germany}
\date{\today}
\begin{abstract}
We show that the $4+1$ dimensional vacuum Einstein equations admit
gravitational waves with radial symmetry. The dynamical degrees of
freedom correspond to deformations of the three-sphere orthogonal
to the $(t,r)$ plane. Gravitational collapse of such waves is
studied numerically and shown to exhibit discretely self-similar
Type II critical behavior at the threshold of black hole
formation.
\end{abstract}

\pacs{Valid PACS appear here}
\maketitle

\section{Introduction and setup} The gravitational collapse to a
black hole is a subject of intensive studies in general
relativity. One of the main goals of these studies is to prove the
cosmic censorship conjecture which says that a physically
realistic generic gravitational collapse cannot result in a naked
singularity. It would be most interesting to assert that this
conjecture is true in vacuum, however with current analytical
techniques the problem seems tractable only in spherical symmetry
and in this case there is no vacuum collapse because of Birkhoff's
theorem. Thus, in order to generate spherically symmetric dynamics
one has to couple matter fields. A simple choice, which led to
important insights, is a real massless scalar field. For this
matter model Christodoulou showed that for small initial data the
fields disperse to infinity \cite{ch1}, while for large data black
holes are formed \cite{ch2}. The transition between these two
scenarios was explored numerically by Choptuik \cite{ch} leading
to the discovery of critical phenomena at the threshold of black
hole formation. Similar phenomena were later observed in many
other matter models in spherical symmetry (see \cite{g} for a
comprehensive review) but, because of numerical difficulties, only
once in vacuum for axially symmetric gravitational waves
\cite{ae}.

 The aim of this letter is to show that - at the price of going to $4+1$ dimensions
  - one can evade Birkhoff's theorem and have
gravitational collapse of pure gravitational waves in radial
symmetry. The idea is very simple and is based on the fact that
the geometry of the three-sphere $S^3$ has the property that one
can break the isotropy but still have a homogeneous space. This
happens as follows. The group of rotations acting on $S^3$ in
Euclidean space has a subgroup $G_3$ acting simply transitively on
the three-sphere. This subgroup is isomorphic to the universal
covering group of the connected  component of the rotation group
in three dimensions. The action of $G_3$ is generated by the
simultaneous rotations in the ($x-y,z-w$)--planes, the
($x-z,y-w$)--planes and ($x-w,z-y$)--planes (this action on $S^3$
defines also the Bianchi IX homogeneous  cosmological model in the
$3+1$ dimensional general relativity). Thus, in $4+1$ dimensions
it is consistent to consider spacetimes with the metric of the
form
\begin{equation}
   \label{as}
ds^2 = -U(t,r)dt^2 +V(t,r) dr^2 + \sum_{k=1}^3 L_k^2(t,r)
\sigma_k^2,
\end{equation}
where $\sigma_k$ are three one-forms invariant under $G_3$
satisfying the commutation relations $d\sigma_i=\frac{1}{2}
\epsilon_{ijk}\,\sigma_j \wedge \sigma_k$.
 In terms of Euler
angles ($0\leq \theta\leq \pi, 0\leq \phi, \psi \leq 2\pi$)
\begin{equation}\label{sigma}
    \sigma_1+i\,\sigma_2=e^{i\psi} (\cos{\theta} \;d\phi + i\,
    d\theta), \quad \sigma_3=d\psi - \sin{\theta}\; d\phi.
\end{equation}
The metric functions $L_k(t,r)$ are the three principal curvature
radii of the squashed three-sphere. The case when all three $L_k$
are equal corresponds to the standard spherically symmetric ansatz
for which the Birkhoff theorem applies and  the only solutions are
Minkowski and Schwarzschild. However, if $L_k$ are different we
will obtain nontrivial vacuum solutions with gravitational
radiation. In this letter we restrict ourselves to a special case
of the ansatz (\ref{as}) in which $L_1=L_2$. Using the coordinate
freedom in the two-space orthogonal to the group orbit of $G_3$,
we choose the volume radial coordinate
$r=(vol(S^3)/2\pi^2)^{1/3}$, and write the metric as
\begin{equation}\label{metric}
    ds^2= - A e^{-2\delta} dt^2 + A^{-1} dr^2 + \frac{1}{4} r^2 \left[ e^{2B}
    (\sigma_1^2+\sigma_2^2) +e^{-4B}\sigma_3^2\right],
\end{equation}
where $A$, $\delta$, and $B$ are functions of $t$ and $r$.  Note
that for this ansatz the three-sphere has a residual isotropy of
the twisted product $S^2 \times S^1$ (as in the Taub universe).

Substituting the ansatz (\ref{metric}) into the vacuum Einstein
equations we get the equations of motion for the functions
$A(t,r), \delta(t,r)$ and $B(t,r)$. In the following we use
overdots and primes to denote $\partial_t$ and $\partial_r$,
respectively. The hamiltonian and momentum constraints are
\begin{eqnarray}\label{mom}
  A'\! &=& \!- \frac{2 A}{r} \!+\frac{1}{3r} \left(8 e^{-2B}\!\!-\!\!2 e^{-8B}\right)\!-\! 2
  r\!
 \left( e^{2\delta} A^{-1} {\dot B}^2\! + \!A {B'}^2\right),\\
    \dot A\! &=& \!- 4 r A \dot B B'.
\end{eqnarray}
The evolution equation for $B$ has a form of the quasilinear wave
equation
\begin{equation}\label{wave}
\left(e^{\delta} A^{-1} r^3 {\dot B}\right)^{\cdot} -
\left(e^{-\delta} A r^3 B'\right)' + \frac{4}{3} e^{-\delta} r
\left(e^{-2B}-e^{-8B}\right)=0.
\end{equation}
  In addition, we have a slicing condition for $\delta$
\begin{equation}\label{rr}
    \delta' = - 2 r \left(e^{2\delta} A^{-2}{\dot B}^2 +
    B'^2\right).
    \end{equation}
It is clear from the above equations that the only dynamical
degree of freedom is the field $B$ which plays a role similar to
the spherically symmetric scalar field in $3+1$ dimensions. If
$B=0$, it is easy to verify (Birkhoff's theorem) that the only
solution is Schwarzschild $\delta_0=0$, $A_0=1-r_h^2/r^2$ (or
Minkowski if $r_h=0$). As we shall see below these two static
solutions play the role of attractors. Note that equations (4-7)
are scale invariant which excludes  existence of regular
asymptotically flat static solutions.

It is convenient to introduce the mass function $m(t,r)$ defined
by $A = 1-m(t,r)/r^2$.
 Then, the hamiltonian constraint (4) takes a simple form
 \begin{equation}\label{mass}
 m' = 2r^3\left(e^{2\delta} A^{-1} {\dot B}^2 + A {B'}^2\right)+\frac{2}{3} r
 \left(3+ e^{-8B}-4 e^{-2B}\right).
\end{equation}
Note that the right hand side of equation (\ref{mass}) is
manifestly positive so $m(t,r)$ is monotone increasing with $r$.
For asymptotically flat spacetimes $m_{\infty}=\lim_{r\rightarrow
\infty}m(t,r)$ exists and is constant in time. The total mass is
given by $M=(3\pi/8G) m_{\infty}$.

We consider the initial value problem for the above equations. To
ensure regularity  at the center we impose the boundary conditions
\begin{equation}\label{bc}
 B(t,r) \sim b(t) r^2, \qquad 1-A(t,r) \sim a(t) r^4.
\end{equation}
We normalize time by the condition $\delta(t,0)=0$, which means
that $t$ is the proper time at the center.
\section{Numerical results} We have solved the above equations
using the free evolution scheme in which $A(t,r)$ is updated using
the momentum constraint (5). The hamiltonian constraint (4) was
only monitored to check the accuracy of the code. The wave
equation (6) was rewritten as the pair of two first order
equations for $B$ and an auxiliary variable $P=e^{\delta}\dot
B/A$. Integration in time was done  by a modified
predictor-corrector McCormack method on a uniform spatial grid.
The ordinary differential equation (\ref{rr}) was solved with the
fourth order Runge-Kutta method. The whole procedure has accuracy
of second order in time and fourth order in space.

The numerical results presented below were produced for  initial
data of the form of an 'ingoing' generalized gaussian
\begin{equation}\label{ic}
    B(0,r)=p\, \left(\frac{r}{r_0}\right)^4
    e^{-(r-r_0)^4/s^4},\quad P(0,r)=r B'(0,r)/r_0,
\end{equation}
where the parameter $p$ was varied and the parameters $r_0$ and
$s$ were fixed. To check universality of the critical behavior, we
have verified that the results are independent of the specific
choice of initial data.

For all families of initial data the same picture, similar to the
massless scalar field collapse, emerges. For small values of the
control parameter $p$ the fields disperse leaving behind the flat
spacetime. For large initial data a Schwarzschild black hole is
formed, as shown in Figs. 1 and 2.

\begin{figure}[h]
\centering
\includegraphics[width=0.46\textwidth]{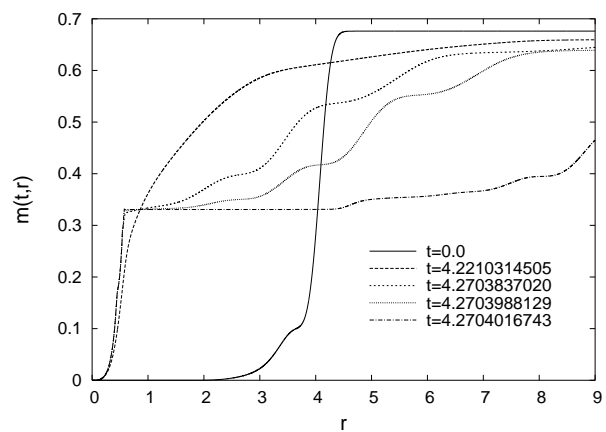}
\caption{\small{Formation of a black hole for highly supercritical
initial data. We plot the mass function $m(t,r)$ at the initial
time and at four late times. The total mass (measured in units
$3\pi/8G$) is $0.67$ (see the plateau of the initial profile).
During the evolution the function $m(t,r)$ develops a second inner
plateau which indicates formation of the Schwarzschild black hole
with mass $M_{BH}=0.33$.} }\label{fig1}
\end{figure}

\begin{figure}[h]
\centering
\includegraphics[width=0.46\textwidth]{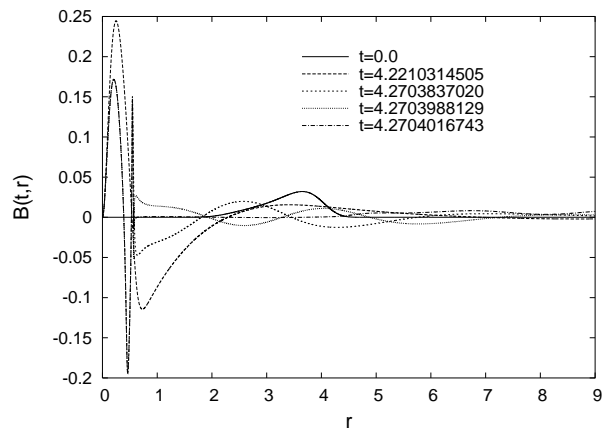}
\caption{\small{The plot of $B(t,r)$ for the same data as in
Fig.1. Outside the horizon developing at $r_h=0.57$, $B$ tends to
zero.} }\label{fig2}
\end{figure}

The process of settling down to the Schwarzschild black hole can
be described in more detail using the linear perturbation theory.
Linearizing  equations (4-7) around the Schwarzschild solution
  we obtain the linear wave equation for the perturbation $\delta
  B(t,r)$
\begin{equation}\label{pert}
\ddot{\delta\!B} - \frac{1}{r^3} A_0 (r^3 A_0
{\delta\!B'})'+\frac{8 A_0}{r^2} \delta\!B=0, \quad
A_0=1-\frac{1}{r^2},
\end{equation}
where we have used the scaling freedom to  set the radius of the
horizon $r_h=1$. Introducing the tortoise coordinate
    $x=r+\frac{1}{2} \ln{\frac{r-1}{r+1}}$
and substituting $\delta B(t,r)=e^{-i k t} u(x)$ into (\ref{pert})
we get the Schr\"odinger equation on the real line
$-\infty<x<\infty$
\begin{equation}\label{schr}
    -\frac{d^2 u}{dx^2} + V(r(x)) u = k^2 u,
    \end{equation}
where
\begin{equation}\label{pot}
    V(r)=\frac{1}{4}
    \left(1-\frac{1}{r^2}\right)\left(\frac{35}{r^2}+\frac{9}{r^4}\right).
\end{equation}
Quasinormal modes (i.e. solutions of equation (\ref{schr})
satisfying the outgoing wave conditions $u\!\sim\! e^{\pm i k x}$
for $x\rightarrow \pm\infty$) for potentials of this type have
been computed in \cite{cly} via the method of continued fractions.
The potential (\ref{pot}) corresponds to the gravitational tensor
perturbation with $l=2$ and in this case the least damped mode
(see Table III in \cite{cly}) has eigenvalue $k=1.51- 0.36\, i$
(in units $r_h^{-1}$). This mode is expected to dominate the
intermediate asymptotics of local convergence to Schwarzschild. In
Fig.~3 we show evidence confirming this expectation.

\begin{figure}[h]
\centering
\includegraphics[width=0.5\textwidth]{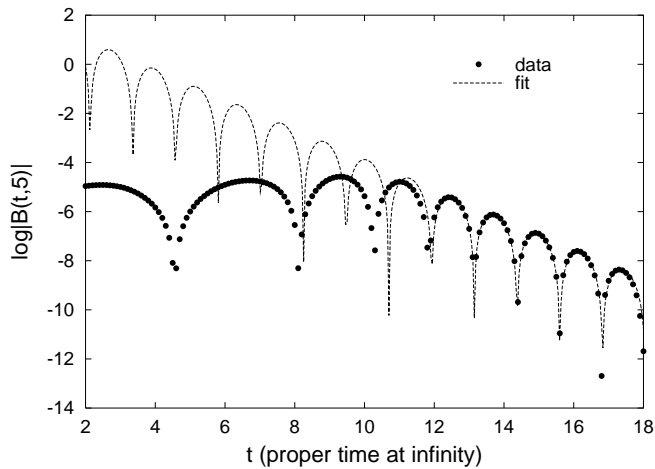}
 \caption{\small{Quasinormal ringing of the Schwarzschild black hole.
 We plot the time series $\ln|B(t,r_0)|$ at $r_0=5$ for the same data as in Fig. 2 but using the proper time at infinity.
 The horizon forms  at $r_h=0.575$, hence according to \cite{cly} the least
 damped quasinormal mode has eigenvalue
$k=2.62-0.62\, i$. The linear
 perturbation regime  sets in at $t\sim 11$.
 The fit of an exponentially  damped sinusoid to the data on the time interval
 $13<t<18$ gives $k=2.56-0.61\, i$, which is in good agreement with the theoretical prediction.} }\label{fig3}
\end{figure}

 We turn now to the description of critical behavior at the threshold of black hole formation.
 In by now routine procedure for bistable systems,
 using bisection we have
determined a critical parameter value $p^*$ which separates black
hole formation from dispersion. The behavior of near-critical
solutions clearly indicates the existence of a discretely
self-similar critical solution with the echoing period $\Delta\sim
0.47$. The profile of a near-critical solution and evidence for
discrete self-similarity are shown in Figs. 4 and 5.

\begin{figure}[h]
\centering
\includegraphics[width=0.5\textwidth]{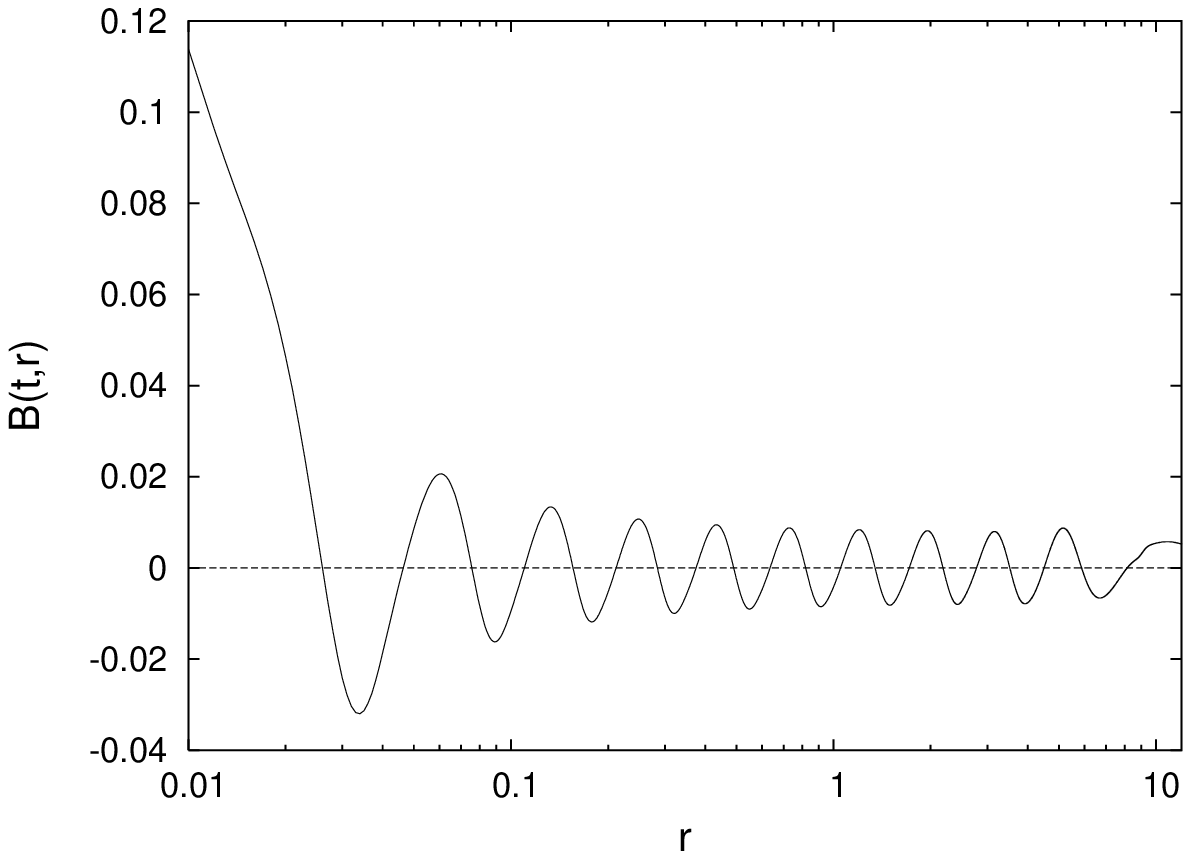}
\caption{\small{The late time profile of $B$ for a near-critical
solution.  We easily get many echoes even for a moderate
fine-tuning because the product of the eigenvalue $\lambda$ of the
growing mode about the critical solution and the echoing period
$\Delta$ is quite small $\lambda \Delta\approx 2.86$ (for
comparison, this product is about three times greater for the
massless scalar field critical collapse \cite{ch}).} }\label{fig3}
\end{figure}
\begin{figure}[h]
\centering
\includegraphics[width=0.5\textwidth]{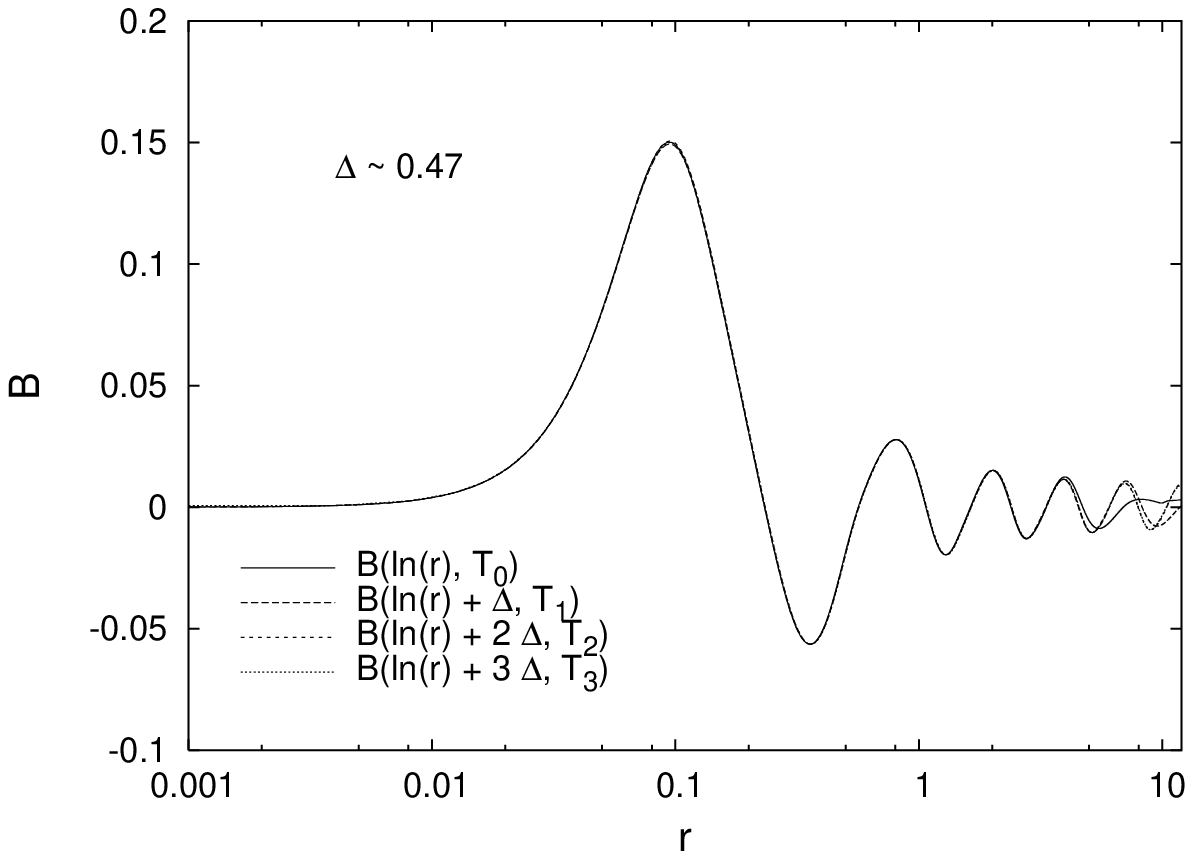}
\caption{\small{Evidence for discrete self-similarity of the
critical solution. For a near-critical evolution we plot the
profile of $B$ at some arbitrary central proper time $T_0$ and
superimpose the next three echoes which subsequently develop. The
times $T_1, T_2, T_3$ and the echoing period $\Delta$ were chosen
to minimize $B(\ln(r)+n\Delta, T_n)-B(\ln(r), T_0)$.}
}\label{fig4}
\end{figure}
\newpage
As expected, the mass of the black hole $M_{BH}(p)$ changes
continuously with $p$ and tends to zero for $p\rightarrow p^*$
according to the power law
\begin{equation}\label{scaling}
    M_{BH} \sim (p^*-p)^{\gamma},
\end{equation}
where the scaling exponent $\gamma \sim 0.3289$ is universal (i.e.
independent of initial data). This is shown in Figs. 6 and 7. Note
that, in four space dimensions, mass has the dimension of
$length^2$, hence $\gamma=2/\lambda$, where $\lambda$ is an
eigenvalue of the growing mode of the critical solution.

\begin{figure}[h]
\centering
\includegraphics[width=0.48\textwidth]{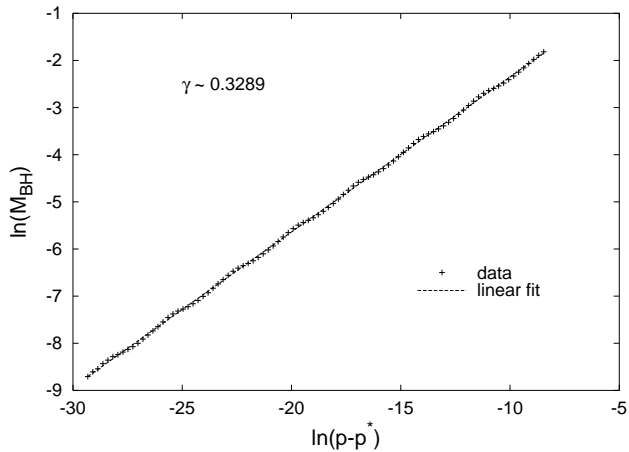}
\caption{\small{Black hole mass scaling. For supercritical
solutions the logarithm of black hole mass $M_{BH}$ (measured in
units $3\pi/8G$) is plotted versus the logarithmic distance to
criticality. The slope of the linear fit yields $\gamma\approx
0.3289$. The wiggles, which are imprints of discrete
self-similarity, are shown in more detail in Fig. 7.}
}\label{fig5}
\end{figure}
\vskip -0.5cm
\begin{figure}[h]
\centering
\includegraphics[width=0.45\textwidth]{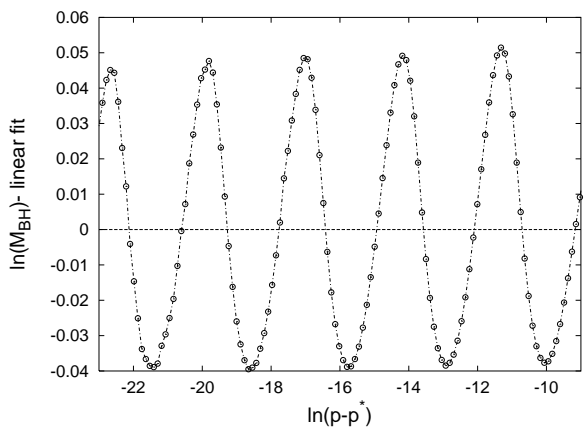}
 \caption{\small{Fine structure of black hole mass scaling. The
linear fit from Fig. 6 is subtracted from the data. The period of
the wiggles agrees to two decimal places with the theoretical
prediction $\Delta/\gamma=1.43$.} }\label{fig6}
\end{figure}

 Summarizing, we have studied gravitational collapse of
pure gravitational waves in $4+1$ dimensions and have found strong
evidence for the Type II discretely self-similar critical behavior
at the threshold of black hole formation. As far as we know,
besides the notable paper \cite{ae}, this is the only example of
critical behavior without matter.

To conclude, let us mention some natural extensions of the study
presented here. One interesting possibility is to investigate the
general ansatz (\ref{as}) with two dynamical degrees of freedom -
the studies in this direction are in progress and will be reported
elsewhere. It is also natural to look for similar models in higher
dimensions. If one insists that the subgroup of the orthogonal
group acts simply transitively, the only possibility is the one
discussed in our paper. However, if one allows multiply transitive
subgroups of the orthogonal groups, the results given in Besse
\cite{besse} show that there are models similar to the one
considered here on any odd-dimensional sphere.
\section{Acknowledgments} We wish to thank Hermann Nicolai for
a helpful discussion. PB acknowledges the friendly hospitality of
the Albert Einstein Institute during part of this work.

\end{document}